\documentclass[12pt]{article}
\usepackage[dvips]{color}
\usepackage{cite,amssymb}
\usepackage{float}
\usepackage{amssymb}
\usepackage{amsmath}
\usepackage{color}
\usepackage[colorlinks=true,linkcolor=red]{hyperref}
\usepackage{hyperref}
\begin{document}
\title{\bf Fermions in the 5D Gravity-Scalar Standing Wave Braneworld}
\author{{\bf Merab Gogberashvili}\\
Andronikashvili Institute of Physics, \\
6 Tamarashvili St., Tbilisi 0177, Georgia \\
and \\
Javakhishvili State University, \\
3 Chavchavadze Ave., Tbilisi 0179, Georgia\\
{\sl E-mail: gogber@gmail.com} \\\\
{\bf Pavle Midodashvili}\\
Ilia State University, \\ 3/5 Kakutsa Cholokashvili Ave., Tbilisi 0162, Georgia\\
{\sl E-mail: pmidodashvili@yahoo.com} \\\\
}
\maketitle
\begin{abstract}
In the article we investigate localization problem for spinor fields within the 5D standing wave braneworld with the bulk real scalar field and show that there exist normalizable fermion field zero modes on the brane.
\vskip 0.3cm
Keywords: Standing waves braneworld; Localization of fermions
\vskip 0.3cm
PACS numbers: 04.50.-h; 11.25.-w; 11.27.+d
\end{abstract}

\vskip 0.5cm

Braneworld models with large extra dimensions \cite{Hi-1,Hi-2,Hi-3,Hi-4,Hi-5,Hi-6} have been very useful in addressing several open questions in high energy physics. Most of the braneworld scenarios considered in literature are realized using time-independent field configurations. However, some models with time-dependent metrics and matter fields also were proposed \cite{S-1,S-2,S-3,S-4}.

Localization problem is the key issue in realization of a braneworld scenario. For reasons of economy and stability one would like to have a universal gravitational trapping mechanism for all fields. Recently we have considered the new class of the non-stationary braneworlds \cite{Wave-1,Wave-2,Wave-3,Wave-4}, which provides the natural localization mechanism by the collective oscillations of the bulk gravitational and scalar fields \cite{Loc-1,Loc-2,Loc-3,Loc-4,Loc-5}.

In this letter we investigate the spinor field localization problem within the 5D standing wave braneworld with a real scalar field, introduced in \cite{Real-1,Real-2}, with the metric {\it ansatz}:
\begin{equation} \label{Metric}
 ds^2 = \frac {e^S}{(1 - a|r|)^{2/3}}\left( dt^2 - dr^2 \right) - (1 - a|r|)^{2/3}\left( e^u dx^2 + e^u dy^2 + e^{-2u}dz^2 \right)~,
\end{equation}
where $a$ is a positive constant. The metric functions $u$ and $S$, which are solutions to 5D Einstein equations, are \cite{Real-1,Real-2}:
\begin{eqnarray} \label{MetricFunctions}
  u(t,|r|)&=& A \sin (\omega t) J_0 \left(\frac{\omega }{a} - \omega|r|\right)~,\nonumber\\
  S(|r|) &=& \frac{3\omega ^2(1 - a|r|)^2}{2a^2}A^2\left[ J_0^2\left( \frac{\omega }{a} - \omega |r| \right) + J_1^2\left( \frac{\omega }{a} - \omega |r| \right) - \right. \\
  &&\left. - \frac{a}{\omega (1 - a|r|)} J_0\left( \frac{\omega }{a} - \omega |r| \right)J_1\left( \frac{\omega }{a} - \omega |r| \right)\right] - B~,\nonumber
\end{eqnarray}
where $A$ and $\omega$ are some constants, $J_0$ and $J_1$ are Bessel functions of the first kind, and the integration constant $B$ is fixed as:
\begin{equation}\label{ConstantB}
  B = \frac 32 A^2\frac{\omega ^2}{a^2}J_1^2\left(\frac{\omega}{a} \right)~.
\end{equation}

We have also the bulk real scalar field standing waves, the solution to the 5D Klein-Gordon equation in the background metric (\ref{Metric}) \cite{Real-1,Real-2},
\begin{equation} \label{BackgroundBulkScalarField}
  \varphi(t,|r|) = \sqrt {\frac{3 M^3}{2}} ~A \cos (\omega t) J_0\left(\frac{\omega }{a} - \omega|r|\right) ~,
\end{equation}
where $M$ denotes the 5D fundamental scale. The scalar field, $\varphi(t,|r|)$, and the metric function, $u(t,|r|)$, are oscillating $\pi/2$ out of phase in time, but have similar dependence on the spatial coordinate $r$.

The metric (\ref{Metric}) has the horizons at
\begin{equation} \label{horizon}
  r = \pm \frac 1a~,
\end{equation}
where some components of the Ricci tensor get infinite values, while all gravitational invariants are finite. This resembles the situation with the Schwarzschild horizon, however, in our case the determinant,
\begin{equation}\label{MetricDeterminant}
  \sqrt g  = (1 - a|r|)^{1/3}e^S~,
\end{equation}
becomes zero at the horizons (\ref{horizon}) and nothing can cross them, i.e. matter fields are confined within the region of size $\sim 1/a$ in the extra space.

Bulk standing gravitational waves $u(t, |r|)$, which is given by (\ref{MetricFunctions}), have the amplitude $A  J_0 \left(\frac{\omega }{a} - \omega|r|\right)$ and the frequency $\omega$. In the extra dimension the waves are bounded by the brane and by the horizons (\ref{horizon}). The standing wave amplitude vanishes at the points where the Bessel function $J_0 (\omega /a - \omega|r|)$ is zero, i.e. these points correspond to nodes of the bulk standing waves. It is also imposed the quantization condition:
\begin{equation}\label{QuantizationCondition}
  \frac{\omega }{a} = Z_N~,
\end{equation}
which guarantees that the brane is located at $r = 0$. In (\ref{QuantizationCondition}) the constant $Z_N$ denotes $N$-th zero of the function $J_0$, and $N \ge 1$ is some fixed integer. So, in the extra dimension, between the horizons (\ref{horizon}), there are $(2N-1)$ nodes at the points:
\begin{equation}\label{AllNodes}
  r_k = \frac{ \operatorname{sgn} (k)}{a} \left( 1 - \frac{Z_{N - \left| k \right|}}{Z_N} \right)~,
\end{equation}
where $k$ runs over $ \pm 1, \pm 2, \cdots ,\pm (N-1)$. One node, we call it $0$-th node, is placed on the brane, i.e. $r_0 = 0$. For simplicity in this paper we assume $N=1$, i.e. standing waves have only one central node located on the brane. In this case for the parameters (\ref{ConstantB}) and (\ref{QuantizationCondition}) we have:
\begin{equation}\label{QuantizationCondition1}
  \frac{\omega }{a} =  Z_1 \approx  2.4~, ~~~~~  B = \frac 32 A^2Z_1^2 J_1^2\left(Z_1 \right)\approx 2.3 A^2~.
\end{equation}

It was shown in  \cite{Real-1,Real-2} that in the case of the large amplitudes of standing waves,
\begin{equation} \label{A}
  A \gg 1~,
\end{equation}
the metric functions $S$ and $u$, which are done by (\ref{MetricFunctions}), obey the relation:
\begin{equation}\label{RatioSu}
  \left| \frac{u}{S} \right| \ll 1~.
\end{equation}
Then the width of the brane,
\begin{equation} \label{d}
  d \sim \frac {1}{aB} \sim \frac {1}{aA^2}~,
\end{equation}
which is located at the origin of the large but finite extra space, $\sim 1/a$, is determined by the metric function $S (|r|)$ \cite{Real-1,Real-2}. So trapping of matter fields on the brane is mainly caused by the pressure of the bulk oscillations and not by the existence of the horizon (\ref{horizon}).

Close to the  0-th node  of the standing wave at $r=0$ the metric (\ref{Metric}) has the approximate form:
\begin{equation} \label{g_AB}
   g_{tt} =  - g_{rr} \approx 1 - \frac rd~,  ~~~~~   g_{xx} =  g_{yy} \approx    g_{zz} \approx - 1 + O\left( \frac {r}{Ad}\right) ~. 
\end{equation}
Then the zero mode wavefunction $\Phi(x^A)$ for a  free brane particle with the energy-momentum
\begin{equation}
P_\nu \ll \frac 1d~,
\end{equation}
can be factorized as:
\begin{equation}
  \left. \Phi \left( x^A \right) \right|_{r \to 0} \approx e^{-i P_\nu  x^\nu } \varphi (r)~,
\end{equation}
where the scalar function $\varphi (r)$ is the extra dimension factor near the brane. So it is obvious from  (\ref{g_AB}) that,  in our case (\ref {A}), the Lorentz violation terms in $g_{xx}$,   $g_{yy}$ and  $g_{zz}$ are much less than KK corrections, $\sim 1/d$, in $g_{tt}$ and $g_{rr}$.

Now let us show existence of the pure gravitational localization of 5D spinor field zero modes on the brane by the background metric (\ref{Metric}) in the case (\ref{A}).

For Minkowskian $4\times 4$ gamma matrices,
\begin{equation}
  \{\gamma^\alpha, \gamma^\beta \} = 2\eta^{\alpha\beta}~,~~~~~~(\alpha, \beta, ...= t,x,y,z)
\end{equation}
we use the representation:
\begin{equation}\label{MinkowskianGammaMatrices}
\begin{array}{l}
\gamma ^t =~ \left( {\begin{array}{*{20}{c}}
I&0\\
0&-I
\end{array}} \right),~~~
{\gamma ^i} = \left( {\begin{array}{*{20}{c}}
0&-\sigma^i\\
\sigma^i&0
\end{array}} \right),~~~
\gamma ^5 = i \gamma^t\gamma^x\gamma^y\gamma^z = \left( {\begin{array}{*{20}{c}}
0&-I\\
-I&0
\end{array}} \right),
\end{array}
\end{equation}
where $I$ and $\sigma^i$ ($i = x,y,z$) denote the standard $2\times2$ unit and Pauli matrices respectively.

Let us recall that fermions in 5D can be represented by four-component columns. The 5D $4 \times 4$ gamma matrices, $\Gamma ^A$, which obey the relations,
\begin{equation}
   \left\{ \Gamma ^A,\Gamma ^B \right\} = 2g^{AB}~,
\end{equation}
can be chosen as:
\begin{equation} \label{Gamma}
  \Gamma^A = h_{\bar A}^A\Gamma^{\bar A}~,~~~~~~\Gamma^{\bar A} = \left(\gamma^t,\gamma^x,\gamma^y,\gamma^z,\gamma^r \right)~,
\end{equation}
where the indices $\bar A,\bar B, ...$, refer to 5D local Lorentz (tangent) frame and  $A$,$B$,... stand for 5D space-time coordinates $t$, $x$, $y$, $z$, $r$,  and
\begin{equation}
\gamma^r=i\gamma^5~.
\end{equation}

The {\it f\"{u}nfbein} for our metric (\ref{Metric}),
\begin{eqnarray}\nonumber
  h_A^{\bar A} &=& \left[ \frac{e^{S/2}}{( 1 - a|r|)^{1/3}}, \frac{(1 - a|r|)^{1/3}}{e^{ - u/2}}, \frac{(1 - a|r|)^{1/3}}{e^{ - u/2}},\frac{(1 - a|r|)^{1/3}}{e^u},\frac{e^{S/2}}{(1 - a|r|)^{1/3}} \right]~, \nonumber \\
  h^{\bar AA} &=& g^{AB}h_B^{\bar A}~, ~~~~~h_{\bar A}^A = \eta _{\bar A\bar B}h^{\bar BA}~, ~~~~~ h_{\bar AA} = \eta _{\bar A\bar B}h_A^{\bar B}~,
\end{eqnarray}
is introduced through the conventional definition:
\begin{equation}\label{VielbeinDefinition}
  g_{AB}=\eta _{\bar A\bar B}h^{\bar A}_A h^{\bar B}_B~.
\end{equation}
According to (\ref{Gamma}), the curved space-time gamma matrices are related to Minkowskian ones by the expressions:
\begin{eqnarray}\label{GammaMatricesRelation}
  \Gamma ^t &=& (1 - a|r|)^{1/3}e^{ - S/2}\gamma ^t~,\\\nonumber
  \Gamma ^x &=& (1 - a|r|)^{-1/3} e^{ - u/2}\gamma ^x~,\\\nonumber
  \Gamma ^y &=& (1 - a|r|)^{-1/3} e^{ - u/2}\gamma ^y~,\\\nonumber
  \Gamma ^z &=& (1 - a|r|)^{-1/3} e^u \gamma ^z~,\\\nonumber
  \Gamma ^r &=& (1 - a|r|)^{1/3}e^{ - S/2}\gamma ^r~.
\end{eqnarray}

The 5D Dirac action for free massless fermions can be written as
\begin{equation}\label{SpinorAction}
  {\cal S} = \int {{d^5}x\sqrt g i\bar \Psi \left( {{x^A}} \right){\Gamma ^M}{D_M}\Psi \left( {{x^A}} \right)}~,
\end{equation}
where the covariant derivatives are:
\begin{equation}\label{CovarianrDerivatives}
  D_A = \partial_A + \frac 14 \Omega_A^{\bar B \bar C} \Gamma_{\bar B} \Gamma_{\bar C}~,
\end{equation}
with $\Omega_M^{\bar M \bar N}$ denoting the spin-connections,
\begin{eqnarray}\label{Spin-Connection}
  \Omega_M^{\bar M \bar N} = - \Omega_M^{\bar N \bar M} = \frac12 \left[ h^{N\bar M}\left( \partial _M h_N^{\bar N} - \partial_N h_M^{\bar N} \right) - h^{N\bar N}\left( \partial_M h_N^{\bar M} - \partial _N h_M^{\bar M} \right) - \right. \nonumber \\
  \left. - h_M^{\bar A} h^{P\bar M}h^{Q\bar N} \left( \partial_P h_{Q\bar A} - \partial_Q h_{P\bar A} \right) \right] ~.
\end{eqnarray}
The non-vanishing components of the spin-connection in the background (\ref{Metric}) are:
\begin{eqnarray}\label{Spin-ConnectionComponents}
  \Omega _t^{\bar t \bar r} &=&  - \frac{S'}{2} - \frac 13 \cdot \frac{a\operatorname{sgn} (r)}{1 - a|r|} ~, \nonumber \\
  \Omega _x^{\bar x \bar r} &=& \Omega _y^{\bar y \bar r} =  - (1 - a|r|)^{2/3}e^{- S/2 + u/2}\left[ \frac{u'}{2} - \frac 13 \cdot \frac{a\operatorname{sgn}(r)}{1 - a|r|} \right]~, \nonumber\\
  \Omega _z^{\bar z \bar r} &=& -(1 - a|r|)^{2/3}e^{- S/2 - u}\left[- u' - \frac{1}{3}\cdot \frac{a \operatorname{sgn}(r)}{1 - a|r|} \right]~,\\
  \Omega _x^{\bar x \bar t} &=& \Omega _y^{\bar y\bar t} = \frac{\dot u}{2}(1 - a|r|)^{2/3}e^{- S/2 + u/2}~,\nonumber\\
  \Omega _z^{\bar z \bar t} &=& - \dot u (1 - a|r|)^{2/3}e^{- S/2 - u}~, \nonumber
\end{eqnarray}
where dots and primes denote derivatives with respect to the time, $t$, and the extra coordinate, $r$, respectively.

Using (\ref{Spin-ConnectionComponents}) for the covariant derivatives (\ref{CovarianrDerivatives}) we find:
\begin{eqnarray}\label{CovarianrDerivativesExplicitForms}
  D_t &=& \partial _t + \left[ \frac{S'}{4} + \frac 16 \cdot \frac{a\operatorname{sgn}(r)}{1 - a|r|} \right]\gamma _r\gamma _t~, \nonumber\\
  D_x &=& \partial _x + (1 - a|r|)^{2/3}e^{- S/2 + u/2}\left\{ \left[ \frac{u'}{4} - \frac 16 \cdot \frac{a\operatorname{sgn}(r)}{1 - a|r|} \right]\gamma _r - \frac{\dot u}{4}\gamma _t \right\}\gamma _x~, \nonumber\\
  D_y &=& \partial _y + (1 - a|r|)^{2/3}e^{- S/2 + u/2}\left\{ \left[ \frac{u'}{4} - \frac 16 \cdot \frac{a\operatorname{sgn}(r)}{1 - a|r|} \right]\gamma _r - \frac{\dot u}{4}\gamma _t \right\}\gamma _y~, \\
  D_z &=& \partial _z + (1 - a|r|)^{2/3}e^{- S/2 - u}\left\{ \left[ - \frac{u'}{2} - \frac 16 \cdot \frac{a\operatorname{sgn}(r)}{1 - a|r|} \right]\gamma _r + \frac{\dot u}{2}\gamma _t \right\}\gamma _z~, \nonumber\\
  D_r &=& \partial _r~. \nonumber
\end{eqnarray}
Then the 5D Dirac equation,
\begin{equation}\label{DiracEquationIn5D}
  i\Gamma^AD_A\Psi = 0~,
\end{equation}
reduces to
\begin{equation}\label{DiracEquationForOurMetric}
  \left\{ \gamma ^t\partial _t + \frac{e^{S/2 - u/2}}{(1 - a|r|)^{2/3}}\left(\gamma ^x\partial _x + \gamma ^y\partial _y + e^{3u/2}\gamma ^z\partial _z \right) + \gamma ^r\left[ \partial _r + \frac{S'}{4} - \frac 13 \cdot \frac{a \operatorname{sgn} (r)}{1 - a|r|} \right] \right\}\Psi  = 0~.
\end{equation}
In the approximation (\ref{RatioSu}) we can ignore the function $u$ in comparison with the function $S$ and this expression gets the form:
\begin{eqnarray}\label{DiracEquationForOurMetric1}
  \left\{ \gamma ^t\partial _t + \frac{e^{S/2}}{(1 - a|r|)^{2/3}}\gamma ^i\partial _i + \gamma ^r\left[ \partial _r + \frac{S'}{4} - \frac 13 \cdot \frac{a \operatorname{sgn} (r)}{1 - a|r|} \right] \right\}\Psi  = 0~,
\end{eqnarray}
where the index $i$ runs over $x$, $y$ and $z$.

In general the solution to (\ref{DiracEquationForOurMetric1}) has non-trivial dependence on 4D and $r$ coordinates. But close to any $k$-th node of standing waves (\ref{AllNodes}) the wavefunction can be factorized as:
\begin{equation}\label{PsiNearKthNode}
  \left. {\Psi \left( {{x^A}} \right)} \right|_{r \to {r_k}} \approx {\psi _k}\left( {{x^\nu }} \right){\rho _k}(r)~,
\end{equation}
where the scalar function $\rho_k (r)$ is the extra dimension factor of the fermion wave function near the $k$-th node. Let us use this decomposition of the spinor wavefunction for two limiting regions: on the brane, where the zeroth node is located,
\begin{equation}\label{PsiOnBrane}
  \left. \Psi \left( x^A\right) \right|_{r \to {0}} \approx \psi _0\left( x^\nu  \right)\rho _0(r)~,
\end{equation}
and at the horizon.
\begin{equation}\label{PsiAtHorizon}
  \left. \Psi \left( x^A\right) \right|_{|r| \to {1/a}} \approx \psi _h\left( x^\nu  \right)\rho _h(r)~.
\end{equation}
\begin{itemize}
\item{On the brane, $r=0$, we assume that 4D part of (\ref{PsiOnBrane}), $\psi _0 \left(x^\nu\right)$, corresponds to the zero mode Dirac spinor, i.e.
\begin{equation}\label{DiracEquationIn4D}
  i\gamma ^\mu \partial _\mu \psi_0 = 0~.
\end{equation}
In this region the metric function $S(r)$ has the following series expansion \cite{Real-1,Real-2}:
\begin{equation}\label{Function-S(r)-AtTheOrigin}
  \left. S \right|_{r \to 0} =  - B\left( a|r| + \frac 12 a^2|r|^2 \right) + O\left( |r|^3 \right)~,
\end{equation}
where the constant $B$ is done by (\ref{ConstantB}). Then it's easy to find that the equation (\ref{DiracEquationForOurMetric1}) reduces to the following asymptotic form:
\begin{equation}\label{DiracEquationIn5DOnBrane}
  \left[ \partial _r - \frac{aB}{4}\operatorname{sgn} (r) \right]\rho_0 (r) = 0~,
\end{equation}
which has the solution:
\begin{equation}\label{RhoOnTneBrane}
  \rho_0 (r) = Ce^{aB|r|/4}~,
\end{equation}
where $C$ is the integration constant.}
\item{Close to the horizons, $|r| \to 1/a$, the metric function $S\left(r\right)$ has the expansion \cite{Real-1,Real-2}:
\begin{equation}\label{Function-S(r)-AtTheHorizon}
  \left. S \right|_{|r| \to 1/a} =  - B\left[1 - \frac{(1 - a|r|)^2}{2J_1^2(\omega /a)} \right] + O\left( (1 - a|r|)^4 \right)~,
\end{equation}
and the equation (\ref{DiracEquationForOurMetric1}) gets the following asymptotic form:
\begin{equation}\label{DiracEquationForOurMetricAtHorizon}
  \left\{ \gamma ^t\partial _t + \frac{e^{ - B}}{(1 - a|r|)^{2/3}}\gamma ^i\partial _i + \gamma ^r\left[ \partial _r - \frac 13 \cdot \frac{a\operatorname{sgn}(r)}{1 - a|r|} \right]\right\}\Psi  = 0 ~.
\end{equation}
Now, using the factorization (\ref{PsiAtHorizon}), and assuming that at the horizon
\begin{equation}
  \psi_h\left(x^{\nu}\right)= const~,
\end{equation}
for the extra dimension factor $\rho_h (r)$ we get the equation:
\begin{equation}\label{DiracEquationIn5DOnHorizon}
  \left[ \partial _r - \frac 13 \cdot \frac{a\operatorname{sgn}(r)}{1 - a|r|} \right]\rho_h (r) = 0~,
\end{equation}
having the solution
\begin{equation}\label{RhoOnTneHorizon}
  \rho_h (r) = D(1 - a|r|)^{- 1/3}~,
\end{equation}
where $D$ is some real constant.}
\end{itemize}

To have a localized field on a brane, its 4D effective Lagrangian, which appears upon integration over extra coordinate of corresponding 5D one, must be non-vanishing and finite. This condition is true for the zero mode spinor field action (\ref{SpinorAction})  in our case. Indeed, the extra dimension space is effectively finite and, in spite that at the horizons the extra dimension part of zero mode (\ref{RhoOnTneHorizon}) is infinitely increasing function, the integral of (\ref{SpinorAction}) over $r$, due to the determinant (\ref{MetricDeterminant}), is convergent. In addition, since the determinant (\ref{MetricDeterminant}) contains the factor $e^S$, and close to the brane $S(r)$ behaves as (\ref{Function-S(r)-AtTheOrigin}) with $B \gg 1$ (see (\ref{QuantizationCondition1}) and (\ref{A})), the main contribution to this integral comes from the region $[-d,+d]$, where $d$ is the effective width of the brane (\ref{d}). The contribution from the rest part of the extra space is exponentially suppressed and is of the order of  $\sim e^{-B}$. So, for the zero mode spinor wavefunction, which close to the brane and at the horizons behaves according to (\ref{PsiOnBrane}) and  (\ref{PsiAtHorizon}), the Dirac action (\ref{SpinorAction}) can be written as:
\begin{eqnarray}\label{SpinorActionForZeroMode}
  {\cal S}^{(0)} &=& \int_{- 1/a}^{+ 1/a} drd^4x  (1 - a|r|)^{1/3}e^S ~i \bar \Psi \left(x^A\right)\Gamma^B D_B \Psi \left(x^A\right) =  \nonumber \\
   &=& \int_{- d}^{+d} dr (1 - a|r|)^{1/3}e^S \rho_0 ^2 \int d^4x ~i\bar\psi_0 (x^\alpha)\gamma ^\mu \partial_\mu \psi_0 (x^\alpha) +  O\left(e^{-B}\right) ~,
\end{eqnarray}
i.e. the zero mode fermion is localized on the brane of the width $d$.

To conclude, in this paper we have explicitly shown the existence of normalizable zero modes of spinor fields (with left or right chirality) on the brane  within the 5D standing wave braneworld with real scalar field. In this short article we don't touch the well known 5D chirality problem (the absence of  a 5D parity operator), which could be solved, for example,  by orbifolding the extra dimension \cite{Chiral-1,Chiral-2}. The chirality problem, as well as another important issue - the stability of the model, will be considered in our future investigations.

\medskip


\noindent {\bf Acknowledgments:} MG was partially supported by the grant of Shota Rustaveli National Science Foundation $\#{\rm DI}/8/6-100/12$. The research of PM was supported by Ilia State University.


\end{document}